\documentclass[twocolumn,floatfix,showpacs,eqsecnum,prb]{revtex4}
\usepackage{graphicx}
\usepackage{amsmath}
\usepackage{amssymb}
\usepackage{bm}
\usepackage{color}

\begin{document}

\title{Dirac boundary condition at the reconstructed zigzag edge of graphene}
\author{J. A. M. van Ostaay}
\affiliation{Instituut-Lorentz, Universiteit Leiden, P.O. Box 9506, 2300 RA Leiden, The Netherlands}
\author{A. R. Akhmerov}
\affiliation{Instituut-Lorentz, Universiteit Leiden, P.O. Box 9506, 2300 RA Leiden, The Netherlands}
\author{C. W. J. Beenakker}
\affiliation{Instituut-Lorentz, Universiteit Leiden, P.O. Box 9506, 2300 RA Leiden, The Netherlands}
\author{M. Wimmer}
\affiliation{Instituut-Lorentz, Universiteit Leiden, P.O. Box 9506, 2300 RA Leiden, The Netherlands}

\date{September 2011}

\begin{abstract}
Edge reconstruction modifies the electronic properties of finite 
graphene samples. We formulate a low-energy theory
of the reconstructed zigzag edge by deriving the modified boundary condition to
the Dirac equation. If the unit cell size of the reconstructed edge is
not a multiple of three with respect to the zigzag unit cell, valleys remain uncoupled
and the edge reconstruction is accounted for by a single angular parameter $\vartheta$. 
Dispersive edge states exist generically, unless $|\vartheta| = \pi/2$.
We compute $\vartheta$ from a microscopic model for the ``reczag'' reconstruction (conversion of
two hexagons into a pentagon-heptagon pair) and show that it can be measured via the local density of states. 
In a magnetic field there appear three distinct edge modes in the lowest Landau level, 
two of which are counterpropagating.
\end{abstract}

\maketitle

\section{Introduction}

The bulk electronic properties of graphene\cite{CastroNeto2009} are modified by edge
effects in a small sample. A prominent example is a narrow ribbon of
graphene which, depending on the exact lattice termination, is either
gapped (semiconducting) or metallic. \cite{Brey2006a} Edge states may form a
flat band which favors spin polarization,\cite{Fujita1996,Nakada1996} and may
have applications in spintronics. \cite{Son2006} Scanning tunneling microscopy
(STM) has provided considerable experimental support for these predicted edge
effects. \cite{Ritter2009,Niimi2006,Kobayashi2006,Tao2011}

\begin{figure}
\includegraphics[width=0.8\linewidth]{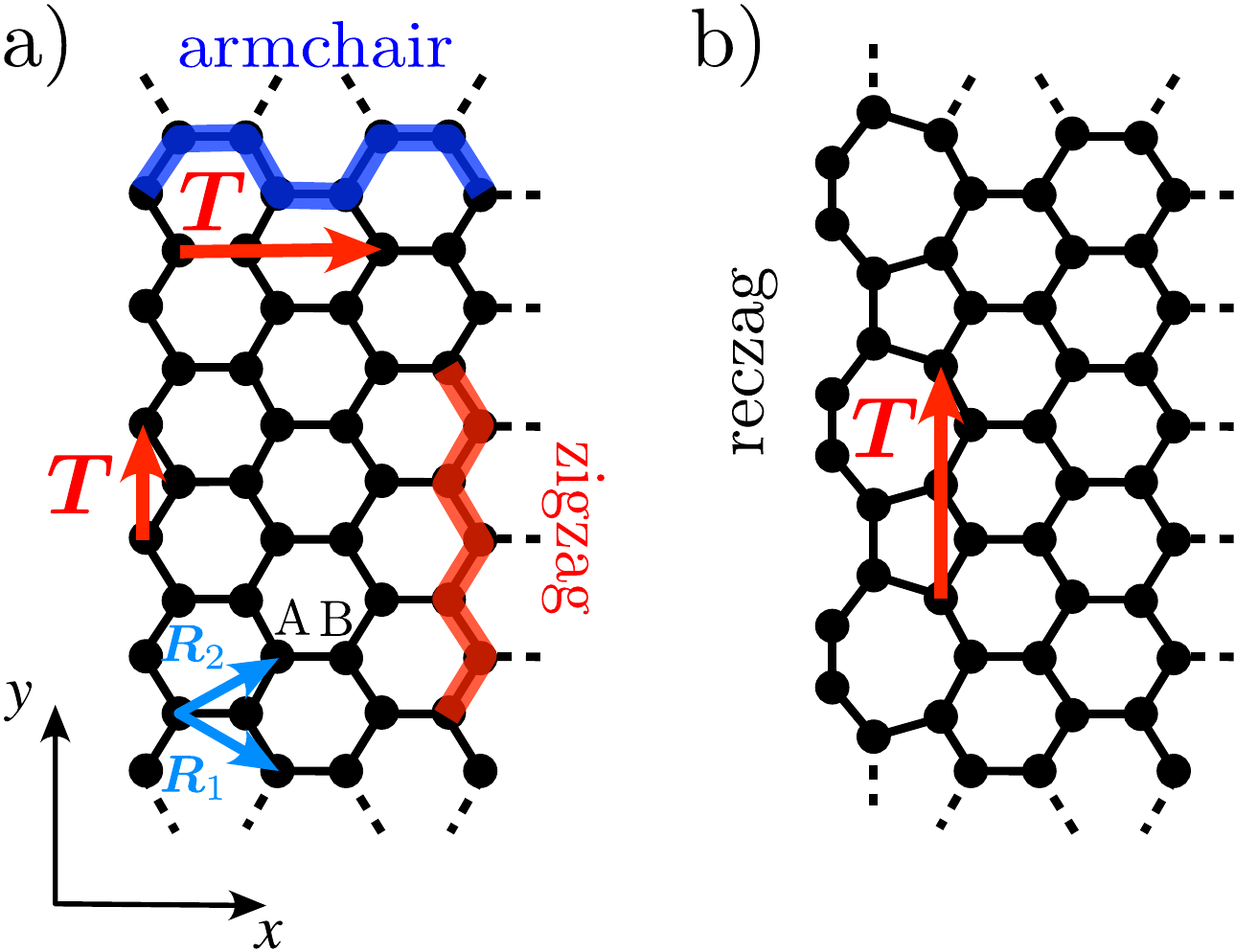}
\caption{(Color online) (a) Zigzag and armchair edge of the honeycomb lattice of graphene. (b) zz(57), or \textit{reczag} reconstruction of the
  zigzag edge.  The translation vector $\bm{T}$ of the various edges is indicated, as well as the Bravais lattice vectors $\bm{R}_1$
  and $\bm{R}_2$ of the honeycomb lattice (with two atoms A and B in the unit cell).}\label{zigzag_reczag}
\end{figure}

The honeycomb lattice of graphene can be terminated along different directions,
with the zigzag and the armchair termination having the smallest unit cell (see
Fig.~\ref{zigzag_reczag}a). Recent microscopic calculations have indicated
that these edges are unstable against a reconstruction of the hexagonal lattice
structure which increases the size of the unit cell. \cite{Koskinen2008,Wassmann2008,
Huang2009,Li2010,Lee2010,Gan2010,Kroes2011} In particular,
Koskinen \textit{et al.}\cite{Koskinen2008}  have shown that the lowest energy is reached for the
zz(57) reconstruction of the zigzag edge: two adjacent hexagons convert
into a pentagon-heptagon pair (see Fig.~\ref{zigzag_reczag}b).
 The stability of this so-called \textit{reczag} edge has
been confirmed by a variety of theoretical calculations
\cite{Wassmann2008,Huang2009,Li2010,Lee2010,Gan2010,Kroes2011} and they have
been observed by transmission electron microscopy. \cite{Koskinen2009,Girit2009}  

Electronic properties of the reczag edge (and related reconstructions) have been studied using the difference
equations obtained from a tight-binding Hamiltonian on the terminated lattice.
\cite{Rakyta2010,Dub10,Rodrigues2011} In this paper we propose an alternative
approach based on the Dirac differential equation,
\cite{McClure1956,DiVincenzo1984} with edge reconstruction accounted for
through a boundary condition. \cite{Akhmerov2008a} The two approaches are
equivalent at low energies, when the wave length is large compared to the
lattice constant. One advantage of the approach based on the Dirac equation is
that it contains fewer independent parameters than the full tight-binding
Hamiltonian. Another advantage is that the
boundary conditions are strongly constrained by symmetry, providing a simple
criterion for the existence of edge states and the presence or absence of
intervalley scattering. 

We show that a broad class of edge reconstructions
can be described by a boundary condition governed by a single angular parameter $\vartheta$. These boundaries cause no intervalley scattering and support dispersive edge states for $|\vartheta|\neq\pi/2$.
The $\vartheta$-class of boundary conditions includes 
any edge reconstruction having a unit cell that is $m$ times the size of
a zigzag unit cell, with $m$ not divisible by three. Most importantly, the 
reczag edge ($m=2$) belongs to the $\vartheta$-class. The value of $\vartheta$ can 
be computed from a microscopic model (and we will carry out this calculation), but we also show how it can be directly measured by STM
via the local density of states.

The paper is organized as follows: In
Sec.~\ref{sec_general_bc}
we begin by discussing the general form of the boundary condition for reconstructed
graphene edges and show how discrete symmetries can be used to 
reduce the number of free parameters to one single parameter (the $\vartheta$-class boundary condition). We then focus in Sec.~\ref{sec_reczag_bc} on the 
particular case of the reczag boundary and compute the numerical value
of $\vartheta$ from a tight-binding model.
Secs.~\ref{sec_nomag} and \ref{sec_mag} are
devoted to a calculation of the electronic structure of graphene terminated by reczag edges
without and with magnetic field, respectively. We conclude
in Sec.~\ref{concl}. The Appendices contain details of the calculations, as well as a discussion of the effects of next-nearest-neighbor hopping and edge potentials on the zigzag boundary condition (which also belongs to the $\vartheta$-class, having $m=1$).

\section{Boundary condition for reconstructed edges}\label{sec_general_bc}

\subsection{Tight-binding and Dirac Hamiltonian}

We describe the electronic structure of graphene using the tight-binding
Hamiltonian,
\begin{equation}
H = \sum_{i,j} t_{ij} \left| i\right> \left<j\right|,
\end{equation}
with one orbital $\left| i\right>$ per atom. In the bulk we restrict ourselves
to uniform nearest-neighbor hopping with value $t$. Only close to the edge we
allow for a reconstruction of the honeycomb lattice and variations in the
hopping amplitudes $t_{ij}$.

In the low-energy limit and sufficiently far from the boundary, 
excitations with energy $\varepsilon$  obey the Dirac equation
\begin{equation}
H\Psi = \varepsilon\Psi, \label{Dirac_equation}
\end{equation}
where the Hamiltonian
\begin{equation}
H = v_{\rm F}\tau_0\otimes \left(\bm{\sigma}\cdot \bm{p}\right)\label{HDiracdef}
\end{equation}
acts on a four-component spinor wave function
\begin{equation}
\Psi =
(\Psi_1,\Psi_2,\Psi_3,\Psi_4)= (\Psi_\text{A},
-i\Psi_\text{B},i\Psi'_\text{B}, -\Psi'_\text{A}).\label{Psidef}
\end{equation}
Here $\Psi_\text{X}$
and $\Psi'_\text{X}$ denote the wave amplitude on the $\text{X}
\in\{\text{A},\text{B}\}$ sublattice in the valley $K$ and $K'$ respectively.
The Fermi velocity is denoted by $v_{\rm F}$ and $\bm{p} = (-i\hbar\partial_x,-i\hbar\partial_y)$ is the two-dimensional
momentum operator. The matrices $\tau_j$ and $\sigma_j$ are the Pauli matrices
in valley and sublattice space respectively (with unit matrices $\tau_0$ and
$\sigma_0$).

The Dirac equation \eqref{Dirac_equation} has a sublattice (or ``chiral'')
symmetry, 
\begin{equation}
(\tau_z \otimes \sigma_z)H (\tau_z\otimes\sigma_z) =- H.
\end{equation}
This symmetry implies that $H \mapsto -H $ for $\Psi_{\text{A}} \mapsto
\Psi_{\text{A}}$ and $\Psi_{\text{B}} \mapsto -\Psi_{\text{B}}$. Physically, it
expresses the fact that the nearest-neighbor hopping does not couple sites on
the same sublattice. Chiral symmetry is preserved by lattice termination, but
it is broken by edge reconstruction (which couples sites originating from the
same sublattice). 

\subsection{Boundary conditions for broken chiral symmetry}

The Dirac equation \eqref{Dirac_equation} must be supplemented by a boundary condition
that also includes the effects of the edge reconstruction. 

In Ref.~\onlinecite{Akhmerov2008a} it was shown that any valid
current-conserving and time-reversally symmetric boundary condition
for the Dirac equation has the form
\begin{equation}
\Psi=M\Psi, \quad M = \left(\bm{\nu}\cdot\bm{\tau}\right) \otimes
\left(\bm{n}\cdot\bm{\sigma}
\right), \quad \bm{n} \perp \bm{n}_B\, ,
\label{dirac_bc}
\end{equation}
where $\bm{n}_B$ is the unit vector in the $x-y$ plane normal to the boundary,
and $\bm{\nu}$ and $\bm{n}$ are three-dimensional unit vectors. If
the edge makes an angle $\alpha$ with the $x$-axis, the boundary condition can
be written more explicitly as 
\begin{equation}\label{general_bc}
\begin{split}
\Psi = (\bm{\nu}\cdot \bm{\tau}) \otimes \bigl(\sigma_z\cos\vartheta  + 
(\sigma_x\cos\alpha+ \sigma_y\sin\alpha)\sin\vartheta
\bigr) \Psi,
\end{split}
\end{equation}
with $\theta\in(-\pi/2,\pi/2]$.

Chiral symmetry requires that
$(\tau_z\otimes\sigma_z)M(\tau_z\otimes\sigma_z) = M$, which restricts the
boundary condition \eqref{dirac_bc} to zigzag ($\bm{\nu} = \pm
\hat{z}$, $\bm{n} = \hat{z}$) or armchair ($\nu_z = n_z = 0$) form. Since edge
reconstruction breaks chiral symmetry, other boundary conditions are allowed.
Still, we can reduce the three independent parameters of the general boundary
condition \eqref{dirac_bc} to a single parameter for a broad class of edge
reconstructions, as we will now show. 

In the following we consider edges that are invariant
under a lattice translation $\bm{T}=n \bm{R}_1 + m \bm{R}_2$, $n,m \in
\mathbb{Z}$, where
$\bm{R}_1 = (\sqrt{3}a/2,-a/2)$ and $\bm{R}_2 = (\sqrt{3}a/2,a/2)$ are the two Bravais 
lattice vectors of graphene. Fig.~\ref{zigzag_reczag} shows the translation vector $\bm{T}$
for the example of the zigzag  edge ($n=-1$, $m=1$), the 
armchair edge ($n=1$, $m=1$) and the reczag edge ($n=-2$, $m=2$).
Due to the translational symmetry the Bloch momentum $k \in
[-\pi/\left|\bm{T}\right|, \pi/\left|\bm{T}\right|]$ along the boundary is a
conserved quantum number. A zone-folding argument, detailed in Appendix 
\ref{app_zonefolding}, shows that
the two Dirac points of graphene project onto the same $k$ if $n=m\text{ mod
}3$ and different $k$ otherwise. Conservation of $k$ then
implies that intervalley scattering is forbidden unless $n=m\text{ mod }3$.

These observations allow for some general statements: Any
reconstruction of the armchair edge has a translational vector $\bm{T}$ 
such that $n=m\text{ mod  }3$, and hence allows for any three-parameter boundary
condition \eqref{general_bc}. In contrast, any reconstruction of the zigzag edge has 
$n=-m$. Hence, if $m$ is not divisible by $3$, the boundary condition does
not mix valleys. In this case $\nu = \pm\hat{z}$ and the boundary condition (for a
given edge orientation $\alpha$) has the single-parameter form
\begin{align}\label{special_bc}
\Psi ={}& \pm\tau_z \otimes \bigl(\sigma_z\cos\vartheta  + 
(\sigma_x\cos\alpha+ \sigma_y\sin\alpha)\sin\vartheta
\bigr) \Psi, \nonumber\\
&\quad\text{if $n\neq m $ mod $3$}.
\end{align}

The reczag boundary has a doubling of the unit cell with respect to zigzag
($m=2$) and hence has boundary condition of the form \eqref{special_bc}. If
however the unit cell is a tripled (or a multiple of a tripled) zigzag unit
cell, the general boundary condition \eqref{general_bc} applies,
\textit{i.e.} valleys are typically mixed. An example of such an edge is the
$Z_{211}$ zigzag reconstruction discussed in Ref.~\onlinecite{Wassmann2008}.

In the remainder of the paper we will focus on the reczag edge, since that has been predicted
to be the most stable reconstruction.\cite{Koskinen2008,Wassmann2008,Huang2009,Li2010,Lee2010,Gan2010,Kroes2011} 
However, we will give most of our results without specifying the angle $\vartheta$, so that they apply to any edge with a boundary condition of the form \eqref{special_bc}.
In order to emphasize this
generality, we consider in Appendix \ref{app_modified_zigzag} a zigzag
edge where chiral symmetry is broken due to edge potentials or next-nearest-neighbor hopping, rather than
due to edge reconstruction.

\section{Boundary condition for the reczag edge}\label{sec_reczag_bc}

\subsection{Tight-binding model}
\label{tbmodel_sec}

\begin{figure}
\includegraphics[width=0.8\linewidth]{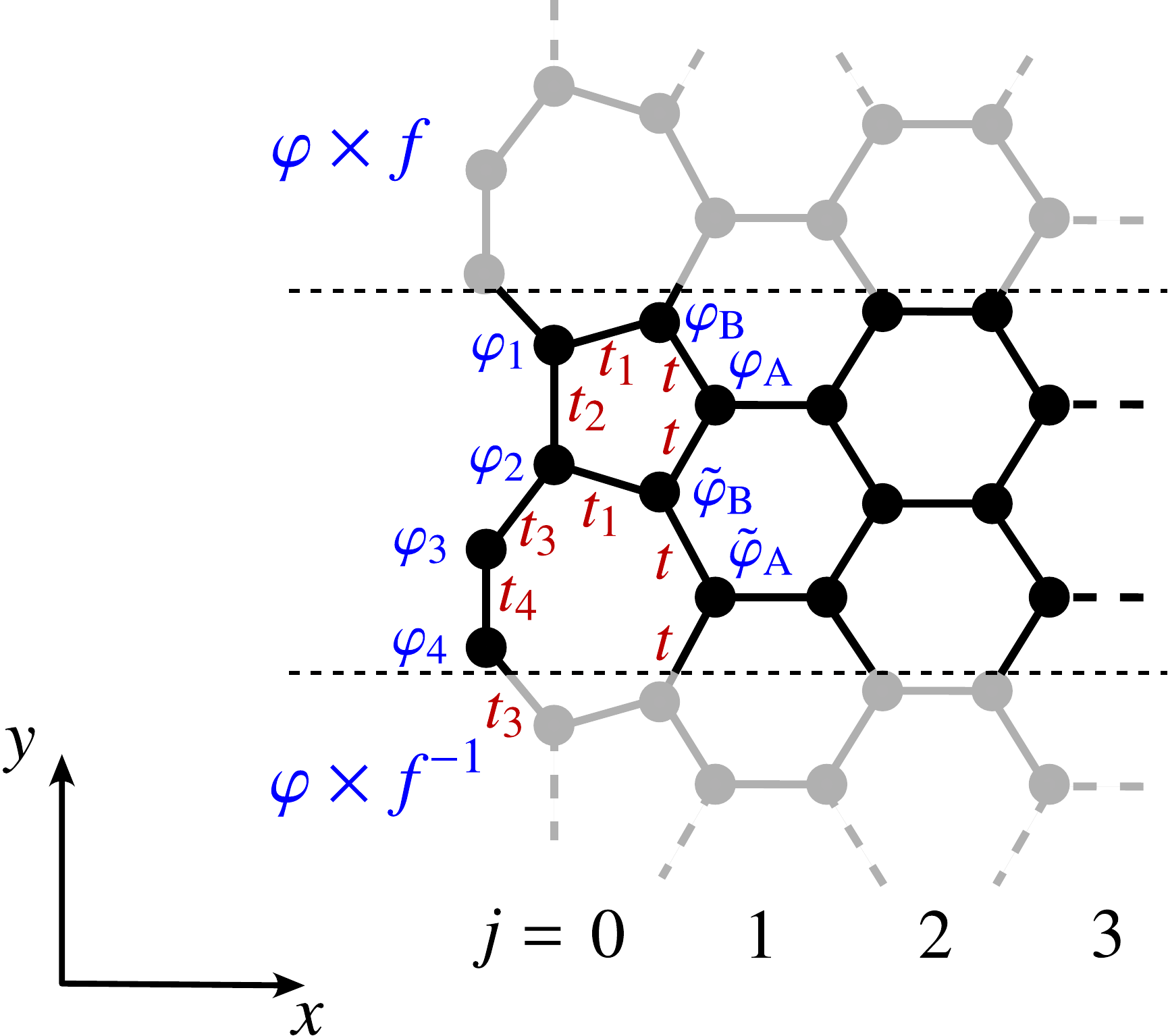}
\caption{(Color online) Nearest-neighbor tight-binding model of the reczag edge with identifiers for the
hopping amplitudes (red) and the wave function amplitudes (blue). We take uniform hopping amplitudes $t$ away from the edge. 
The unit cell of the reczag edge is indicated
in dark, the neighboring unit cells in light color. The wave functions in the
neighboring unit cells are multiplied by a Bloch phase factor $f=e^{2ika}$.
}\label{reczag_tb}
\end{figure}

In order to obtain a value for the angle $\vartheta$ in Eq.~\eqref{special_bc}
for the reczag edge, we employ a tight-binding parametrization.
We consider a reczag edge parallel to the $y$-axis ($\alpha=90^\circ$), as
shown in Fig.~\ref{reczag_tb}. The unreconstructed edge would have terminated
with an atom of the B-sublattice and we will therefore refer to the edge
as the B-type reczag. (We give results for the A-type reczag at the end of
the section.) The boundary condition for a B-type reczag edge along the $y$-axis reads
\begin{equation}
\Psi=-\tau_z \otimes (\sigma_z \cos\vartheta + \sigma_y \sin\vartheta) \Psi.\label{bc_KB}
\end{equation}
We may write this boundary condition more explicitly in terms of the sublattice amplitudes \eqref{Psidef},
\begin{align}
\label{bc_isotropic_B}
&\Psi_1=i \mathcal{F} \Psi_2,\quad \Psi_3=-i\mathcal{F}^{-1} \Psi_4,\\
&\Psi_{\rm A} = \mathcal{F} \Psi_{\rm B},\quad \Psi'_{\rm A} = \mathcal{F} \Psi'_{\rm B},\label{bc_AB}
\end{align}
with the definition
\begin{equation}
{\cal F}=\tan(\vartheta/2).
\end{equation}

The reczag edge is translationally invariant over a distance $2a$, where $a$ is the graphene lattice
constant. Hence, wave functions in adjacent unit cells only differ
by a phase $f=e^{2ika}$, with Bloch wave vector $k$. We allow
for a variation of the hopping amplitude due to the reconstruction, but assume
for simplicity that the hopping amplitude on every hexagon remains given by
the bulk value $t$.

Numerical values for the modified hopping amplitudes from density
functional theory (DFT) are in the literature\cite{Rakyta2010} (see Table \ref{tb_params}). An extended
model for the reczag edge with more parameters has been studied in Ref.~\onlinecite{Rodrigues2011}.
We give results for the extended model in Appendix \ref{app_extended_reczag} and
show that there are no essential differences to the simpler model employed here.
We also neglect the effects of hoppings beyond nearest-neighbor and edge potentials. These effects
can all be accounted for by a modification of the numerical value of $\vartheta$ (see Appendices \ref{app_modified_zigzag} and \ref{app_extended_reczag}).
\begin{table}[!tb]
\begin{tabular}{|c|c|c|c|c|c|}
\hline
$t_1/t$&$t_2/t$&$t_3/t$&$t_4/t$&${\cal F}$&$\vartheta$\\
\hline\hline
$0.91$&$0.99$&$0.97$&$1.5$&0.0753&0.150\\
\hline
\end{tabular}
\caption{DFT values for the hopping amplitudes $t_{p}$ in the tight-binding model
for the reczag edge, from Ref.\ \onlinecite{Rakyta2010}, and the corresponding value of the boundary condition parameter ${\cal F}=\tan(\vartheta/2)$, calculated from Eq.\ \eqref{F_value}.}
\label{tb_params}
\end{table}

Labeling wave function and
hopping amplitudes as indicated in Fig.~\ref{reczag_tb}, we can write down
the tight-binding equations,
\begin{subequations}
\label{edge_equations}
\begin{align}
\varepsilon\varphi_B &= t_1\varphi_1 + t\left(\varphi_A +
f\tilde{\varphi}_A\right), \\
\varepsilon \tilde{\varphi}_B &= t_1\varphi_2 + t\left(\varphi_A +
\tilde{\varphi}_A\right), \\
\varepsilon\varphi_1 &= t_1\varphi_B + t_2\varphi_2 + ft_3\varphi_4, \\
\varepsilon\varphi_2 &= t_2\varphi_1 + t_3\varphi_3 + t_1\tilde{\varphi}_B, \\
\varepsilon\varphi_3 &= t_3\varphi_2 + t_4\varphi_4 , \\
\varepsilon\varphi_4 &= t_3\varphi_1/f + t_4\varphi_3.
\end{align}
\end{subequations}
In the limit $\varepsilon \rightarrow 0$ it is now straightforward to find relations
for the wave functions on the first hexagons away from the reconstructed edge,
\begin{subequations}
\label{intermediate_results}
\begin{align}
\varphi_A &=
\frac{ft^2_1t_4}{(1-f)t}\left[\frac{\varphi_B}{ft_3^2-t_2t_4} -
\frac{\tilde{\varphi}_B}{t_3^2-ft_2t_4} \right], \\
\tilde{\varphi}_A &= \frac{t^2_1t_4}{(1-f)t}
\left[\frac{\tilde{\varphi}_B}{f^{-1}t_3^2 - t_2t_4} -
\frac{\varphi_B}{ft_3^2-t_2t_4} \right].
\end{align}
\end{subequations}

\subsection{Boundary modes}

We proceed along the lines of Ref.~\onlinecite{Akhmerov2008a}, by separating the wave function $\psi$ into a
part $\Psi$ that obeys the Dirac equation, plus a boundary correction
$\psi_{\text{bdy}}(\bm{r})$. Since the valleys are not coupled,
it is sufficient to consider a single valley at $\bm{K}=(0,K)=(0, -4\pi/3a)$,
\begin{subequations}
\begin{align}
\psi_\text{A}(\bm{r}) &= \Psi_\text{A}(\bm{r}) e^{i\bm{K}\cdot \bm{r}}
+ \psi^\text{A}_{\text{bdy}}(\bm{r}) ,\\
\psi_\text{B}(\bm{r}) &= \Psi_\text{B}(\bm{r})e^{i\bm{K}\cdot \bm{r}} 
+ \psi^\text{B}_{\text{bdy}}(\bm{r}).
\end{align}
\end{subequations}

Taking further into account the translational symmetry along the
$y$-direction we can write the wave function as 
\begin{subequations}
\label{wave_functions}
\begin{align}
&\psi_\text{A}(\bm{r}) = \phi_\text{A}(j)e^{iKy}  + \phi^\text{A}_{\text{bdy}}(j)e^{i\tilde{K}y},\\
&\psi_\text{B}(\bm{r}) = \phi_\text{B}(j)e^{iKy} + \phi^\text{B}_{\text{bdy}}(j)e^{i\tilde{K}y},\\
&\tilde{K}=K+\pi/a=-\pi/3a.
\end{align}
\end{subequations}
The index $j$ numbers the unit cells transverse to the edge,
with $\varphi_B$, $\tilde{\varphi}_B$ corresponding to $j=0$
and $\varphi_\text{A}$, $\tilde{\varphi}_\text{A}$ to $j=1$ (see Fig.~\ref{reczag_tb}). 
We denote by $\tilde{K}$ the projection of the $K$-point
into the doubled unit cell of the reczag edge. The Dirac modes thus have a periodicity
given by the unperturbed graphene lattice, whereas the boundary modes are governed by
the periodicity of the reczag reconstruction. 
Application of the boundary condition \eqref{bc_KB} on the Dirac modes specifies the angle 
$\vartheta\in(-\pi/2,\pi/2]$ from
\begin{equation}
\phi_\text{A}(0)/\phi_\text{B}(0) = \tan(\vartheta/2).
\label{boundary_condition_2}
\end{equation}

For the bulk of graphene away from the edge, the tight-binding equations
take the form
\begin{subequations}
\label{graphene_equations}
\begin{align}
\varepsilon\psi_\text{A}(\bm{r})& = t\left[\psi_\text{B}(\bm{r}) + \psi_\text{B}(\bm{r}-
\bm{R}_1) + \psi_\text{B}(\bm{r} - \bm{R}_2) \right], \\
\varepsilon\psi_\text{B}(\bm{r})&= t\left[\psi_\text{A}(\bm{r}) + \psi_\text{A}(\bm{r} +
\bm{R}_1) + \psi_\text{A}(\bm{r}+ \bm{R}_2)\right].
\end{align}
\end{subequations}
Inserting the decomposition \eqref{wave_functions} into Eq.~\eqref{graphene_equations}
and accounting for the fact that the Dirac and boundary modes have a different 
periodicity, we arrive in the limit $\varepsilon \rightarrow 0$ at
\begin{subequations}
\label{connection_with_bulk}
\begin{align}
\phi_\text{A}(j + 1) & = \phi_\text{A}(j),\;\;
\phi^\text{A}_{\text{bdy}}(j + 1)  = \frac{1}{\sqrt{3}}\phi^\text{A}_{\text{bdy}}(j), \\
\phi_\text{B}(j + 1)& = \phi_\text{B}(j) , \;\;
\phi^\text{B}_{\text{bdy}}(j + 1)  = \sqrt{3}\phi^\text{B}_{\text{bdy}}(j).
\end{align}
\end{subequations}
In order for the wave function to be normalizable only non-growing 
contributions are allowed, so $\phi^\text{B}_{\text{bdy}}(j)=0$ for all $j$. The
B-type reczag edge thus has a boundary mode on the A sublattice only. This
boundary mode is a direct consequence of the unit cell doubling of the
reconstructed edge.

The boundary mode decays exponentially away from the edge, with a decay length of 
$3a/2$. This is also the distance from the edge where the Dirac
equation --- which does not capture the boundary modes --- is valid. Hence, the
reczag edge can be faithfully treated within the Dirac approach, as there
are deviations only within the first few unit cells away from the
boundary.

\subsection{Boundary condition}

The wave amplitudes $\varphi_\text{A,B}$ and $\tilde{\varphi}_\text{A,B}$
near the reczag edge can be written in terms of the Dirac and boundary modes as
\begin{subequations}
\begin{align}
\varphi_\text{A} &= \left[\phi_\text{A}(1) + \phi^\text{A}_{\text{bdy}}(1)\right]f^{-1/4} , \\
\tilde{\varphi}_\text{A} &= \left[\phi_\text{A}(1) - \phi^\text{A}_{\text{bdy}}(1) \right]f^{-3/4},\\
\varphi_\text{B} &= \phi_\text{B}(0) ,\\
\tilde{\varphi}_\text{B} &= \phi_\text{B}(0)f^{-1/2}.
\end{align}
\label{boundarymodes}
\end{subequations}
With this decomposition we find from Eq.~\eqref{intermediate_results} that
\begin{align}
&\phi_\text{A}(0)  = \mathcal{F}\, \phi_\text{B}(0),\label{A_Dirac_boundary}\\
&\mathcal{F}=\tan(\vartheta/2)=\frac{t_1^2t_4\left(t_2t_4 - t_3^2\right)}{2t\left(t_3^4 +
t_2t_3^2t_4 + t_2^2t_4^2\right)}.\label{F_value}
\end{align}
The numerical values for ${\cal F}$ and $\vartheta$ for the reczag edge are given in Table \ref{tb_params}.

This concludes the derivation of the boundary condition for the B-type reczag edge. For the A-type reczag, the
role of the A and B sublattices is interchanged. We thus have the boundary conditions
\begin{align}
\label{bc_isotropic_A}
&\Psi_1=i\mathcal{F}^{-1} \Psi_2,\quad \Psi_3=-i\mathcal{F} \Psi_4,\\
&\Psi_{\rm B} = \mathcal{F} \Psi_{\rm A},\quad \Psi'_{\rm B} = \mathcal{F} \Psi'_{\rm A},
\end{align}
with the same value \eqref{F_value} of ${\cal F}$.
  
The zigzag boundary condition\cite{Brey2006a} corresponds to $\mathcal{F}=0$ or $\mathcal{F} = \infty$. 
As one can see from Eq.~\eqref{F_value}, $\mathcal{F}$ vanishes if
$t_3=\sqrt{t_2 t_4}$, so for these matched hopping amplitudes the doubling of
the unit cell at the edge has no effect on the boundary condition. This explains
why a zigzag-edge behavior was found in a tight-binding study of edge
reconstruction for the special case that all hopping amplitudes have their bulk
values.\cite{Rodrigues2011}  

\section{Electronic states}\label{sec_nomag}

\subsection{Dirac solutions}

The knowledge of the boundary condition allows us to calculate
electronic properties.
In this section we consider zero magnetic field and
then in the next section the effect of a magnetic field is included.
Although we use the numerical values of the reczag edge
obtained in the previous Section for plots and comparisons to tight-binding
models, the analytical results we obtain are valid for arbitrary angles
$\vartheta$.

Since the reczag edge does not mix the valleys, it is possible to
consider the $K$ and $K'$-points separately. From Eqs.~\eqref{bc_isotropic_B}
and \eqref{bc_isotropic_A} we see that, given a solution for a particular
valley, substitution of $\mathcal{F} \rightarrow -1/\mathcal{F}$ gives a solution
in the other valley. In what follows we focus our discussion on the
$K$-point.

We consider either one or two reczag edges along the $y$-direction.
The solution of the Dirac equation \eqref{HDiracdef} at energy $\varepsilon$ has
the form $\Psi(x,y)=\psi(x) e^{i k y}$ with
\begin{equation}\label{ansatz_dirac}
\psi(x)=
A \begin{pmatrix}
\frac{\hbar v_\text{F}}{\varepsilon} (k+iq)\\
i
\end{pmatrix} e^{iqx}
+B \begin{pmatrix}
\frac{\hbar v_\text{F}}{\varepsilon} (k-iq)\\
i
\end{pmatrix} e^{-iqx}.
\end{equation}
The wave vector $k$ is real, $q$ is real or imaginary, and the dispersion relation is $\varepsilon=\pm \hbar v_\text{F} \sqrt{k^2+q^2}$. 
The relative amplitudes $A,B$ of the superposition have to be determined by the boundary condition.

\subsection{Edge state dispersion}

To study the dispersion relation of the edge state we take a semi-infinite
graphene sheet for $x \geq 0$, terminated with a B-type reczag
edge at $x=0$.

We first focus on decaying solutions with an imaginary $q=i z$
and energy $|\varepsilon| < |\hbar v_\text{F} k|$. These 
edge states are affected most prominently
by the edge reconstruction. Keeping only the exponentially
decaying part of \eqref{ansatz_dirac} and substituting 
the boundary condition \eqref{bc_isotropic_B}, we find the equation
\begin{equation}\label{edge_state_cond}
\hbar v_\text{F} (z-k) = \mathcal{F} \varepsilon.
\end{equation}
This only has a normalizable solution for
\begin{equation}
z=k 
\frac{1-\mathcal{F}^2}{1+\mathcal{F}^2}=k\cos\vartheta > 0.
\end{equation}
The normalized edge state wave function then reads
\begin{equation}
\psi^\text{edge}(x)=\begin{pmatrix}
i \sin^2 \vartheta/2\\
\cos^2 \vartheta/2
\end{pmatrix}
\sqrt{2 k \cos\vartheta} \,e^{-kx\cos\vartheta} 
\end{equation}
with energy\cite{Akhmerov2008a,Volkov2009}
\begin{equation}\label{edge_disp1}
\varepsilon(k)=-\hbar v_\text{F}\, k \sin\vartheta,\;\; \text{for $k\cos\vartheta>0$.}
\end{equation}

The solution for the $K'$-valley is found by the replacement of $\mathcal{F}
\rightarrow -1/\mathcal{F}$ in Eq.~\eqref{edge_state_cond}, yielding a solution
with energy
\begin{equation}
\varepsilon(k)=\hbar v_\text{F}\, k \sin\vartheta,\;\; \text{for $k\cos\vartheta<0$.}
\end{equation}
These edge states exist for any $|\vartheta|<\pi/2$.

It is instructive to compare the reczag edge state with the well-known zigzag
counterpart,\cite{Fujita1996,Brey2006a} which corresponds to the limit
$\vartheta\rightarrow 0$. In accord with the tight-binding calculations,\cite{Rodrigues2011} 
the main difference between the two types of edge states is their
energy dispersion: While the zigzag edge state features a dispersionless band
$\varepsilon(k)=0$, the reczag edge state has a linear dispersion with 
velocity $v_\text{F} \sin\vartheta$. This has implications for the density of
states (see Sec.\ \ref{sec:DOS}). 

Furthermore, the zigzag edge state is exactly zero
on one sublattice (the A sublattice for a B-type zigzag edge), whereas the
reczag edge couples the two sublattices.
The coupling is such that
the two components of the wave function only differ by a constant factor,
$\psi_1(x)=\mathcal{F}\psi_2(x)$ for all $x$, not only at the boundary. The
wave function thus has \textit{the same} decay length $(k\cos\vartheta)^{-1}$ into the bulk on each sublattice.\cite{note1}
For $\vartheta\rightarrow\pi/2$ the decay length diverges and the edge state disappears in the bulk.

\begin{figure}
\includegraphics[width=0.95\linewidth]{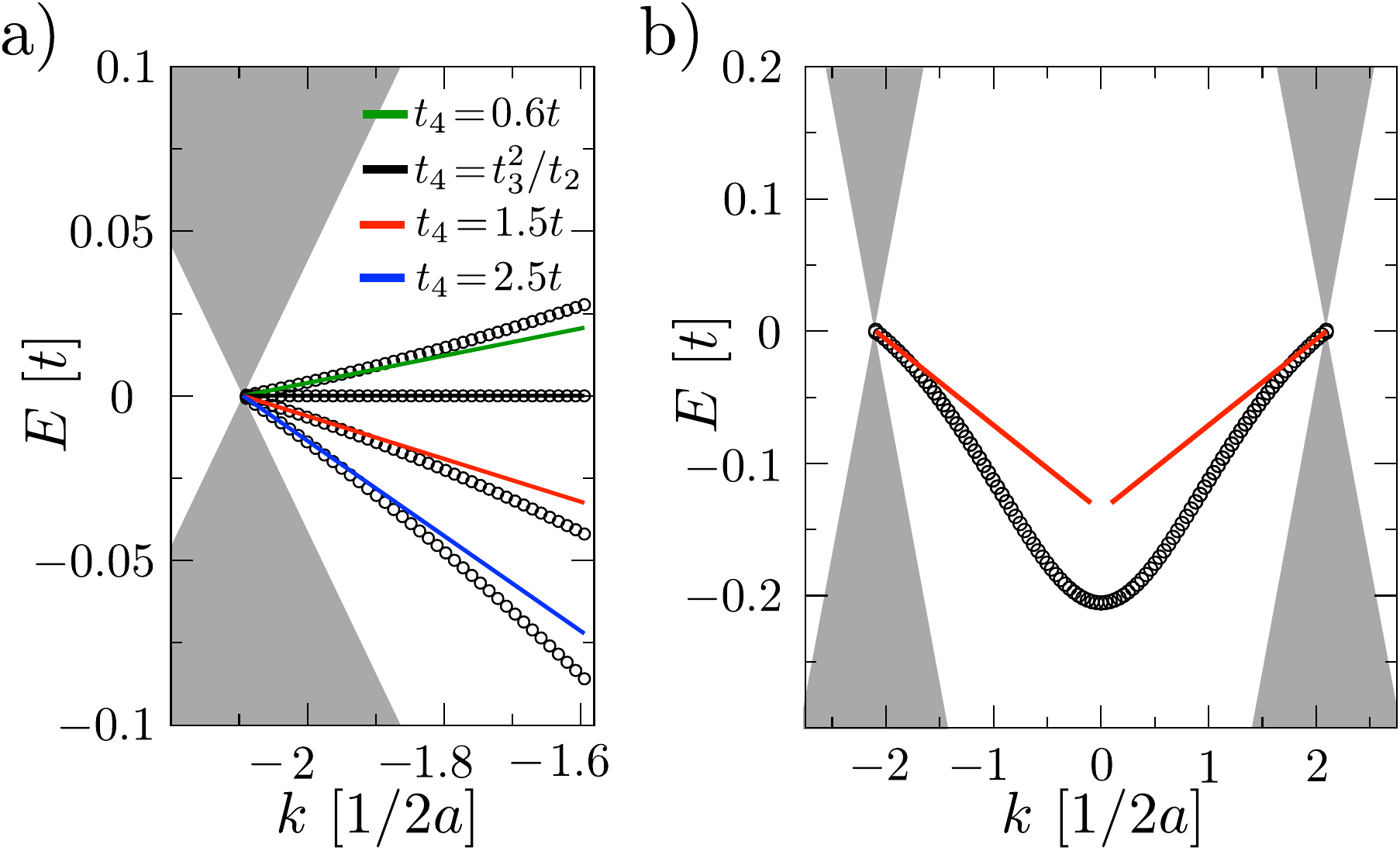}
\caption{(Color online) (a) Comparison between tight-binding (black circle) and Dirac equation
results (solid lines) for the edge state dispersion near the $K$-point. Results are shown for different values
of the hopping amplitude $t_4$ of the reczag edge. (All other hopping amplitudes are as in Table \ref{tb_params}.) The continuum of bulk states in the Dirac cone
is indicated in grey.
(b) The data for $t_{4}=1.5t$ on a larger scale, showing both the $K$ and $K'$ points.
}\label{comp_tb_dirac}
\end{figure}

In Fig.~\ref{comp_tb_dirac}a we compare the edge state dispersion
\eqref{edge_disp1} with the results of the tight-binding model of the reczag edge. (The tight-binding results were calculated for a nanoribbon of width $W=1000 \sqrt{3} a$, large enough that
the opposite edges were essentially decoupled.) Results are shown
for different values of $\mathcal{F}$, obtained by modifying the value of $t_4$ with respect to
the DFT values in Table \ref{tb_params}. As expected, we find excellent agreement for small $\varepsilon$, corresponding to $k$-values close to the $K$ or $K'$ points.
Away from these Dirac points, the two disconnected edge states of the Dirac equation are connected by the tight-binding model, see Fig.~\ref{comp_tb_dirac}b.

\subsection{Density of states}
\label{sec:DOS}

To make contact with STM experiments, we calculate the local density of states (DOS)
on sublattice $j=\text{A,B}$, given by
\begin{equation}
\mathcal{D}_j(\varepsilon, \bm{r})= \sum_n \delta(\varepsilon - \varepsilon_n) \left(
\left | (\Psi_n)_j(\bm{r})\right|^2 +  \left |( \Psi'_n)_j(\bm{r})\right|^2\right).
\end{equation}
The sum runs over all eigenstates $\Psi_n,\Psi'_n$ in valley $K,K'$ with energy $\varepsilon_n$.
For the reczag edge state we find
\begin{subequations}
\begin{align}
\mathcal{D}^\text{edge}_\text{A}(\varepsilon, x)& =
\mathcal{F}^2\,\mathcal{D}^\text{edge}_\text{B}(\varepsilon, x),\\
\mathcal{D}^\text{edge}_\text{B}(\varepsilon, x)& = \frac{g_\text{s}
g_\text{v} \cos^2(\vartheta/2)}{\pi \hbar
v_\text{F}\left|\sin\vartheta\right| } u \Theta(u)e^{- 2 u x},\nonumber\\
& {\rm with}\;\;u=-\varepsilon(\hbar v_{\rm F}\tan\vartheta)^{-1}.
\end{align}
\end{subequations}
The function $\Theta(u)$ is the unit step function [$\Theta(u)=1$ for $u\geq 0$ and zero otherwise]. The coefficients $g_\text{s}=g_\text{v}=2$ indicate the degeneracies due to the spin and valley degree of freedom.
 
Integrating out the transverse coordinate $x$ and summing over both sublattices we find the total
DOS per unit length of the edge,
\begin{equation}
 \mathcal{D}^\text{edge}(\varepsilon)=\frac{g_\text{s} g_\text{v}}{2 \pi
\hbar v_\text{F} \left|\sin\vartheta\right|}\,
 \Theta(u).
\end{equation}
This result holds in the energy range $|\varepsilon|\lesssim (\hbar v_\text{F}/a) |\sin\vartheta|$ (beyond which the Dirac equation breaks down). Such a constant
DOS was also found for the case that the edge state
acquires a dispersion due to next-nearest neighbor hopping.\cite{Wimmer2010,Wurm2011}
Compared to the zigzag case, where $\mathcal{D}^\text{edge}(\varepsilon) \sim \delta(\varepsilon)$,
the density of states is greatly reduced by the reconstruction, which may well prevent the
ferromagnetic instability of the zigzag edge.\cite{Fujita1996}

In addition to the decaying edge state with imaginary $q$, there is a continuum of bulk states with real $q$. Then the term
\begin{equation}
\frac{\hbar v_\text{F}}{\varepsilon} (k+iq)=\text{sgn}(\varepsilon) \frac{k+iq}{\sqrt{k^2+q^2}}
=\text{sgn}(\varepsilon)\, e^{i\varphi}
\end{equation}
in Eq.~\eqref{ansatz_dirac} is a pure phase ($\text{sgn}$ is
the sign function). These bulk solutions are given by
\begin{equation}
\psi^\text{bulk}(x)= C_\text{bulk} \begin{pmatrix}
\sin qx+\text{sgn}(\varepsilon)\, \mathcal{F} \sin(qx+\varphi)\\
i\,\text{sgn}(\varepsilon) \sin(qx-\varphi)+i\mathcal{F} \sin qx
\end{pmatrix}\,,
\end{equation}
with $\hbar v_{\rm F}q=|\varepsilon|\sin\varphi>0$ and normalization constant
\begin{equation}
C_\text{bulk}=\pi^{-1/2}\big(1 + \mathcal{F}^2 +2\,\text{sgn}(\varepsilon)\,\mathcal{F}\cos\varphi\bigr)^{-1/2}\,.
\end{equation}

The local DOS of the bulk states follows upon integration,
\begin{equation}\label{dos_bulk_exact}
\mathcal{D}^\text{bulk}_j(\varepsilon, x) = \frac{g_\text{s} g_\text{v}\left|\varepsilon\right|}{2\pi\hbar^2 v_\text{F}^2}
\int_{0}^\pi d \varphi \left| \psi^\text{bulk}_j(x) \right|^2.
\end{equation}
For ${\cal F}\ll 1$ the integral can be evaluated analytically,
\begin{subequations}
\begin{align}
&\mathcal{D}^{\text{bulk}}_{\rm A}(x, \varepsilon) =
\frac{g_\text{s} g_\text{v}|\varepsilon|}{4\pi\hbar^2 v^2_F}\biggl(1-J_0(\xi) + 2\,\text{sgn}(\varepsilon)\mathcal{F}J_1(\xi)
\nonumber \\
&\qquad + \mathcal{F}^2\bigl[J_0(\xi) - J_2(\xi)\bigr] +{\cal O}({\cal F}^{3}) \biggr),  \\
&\mathcal{D}^{\text{bulk}}_{\rm B}(x, \varepsilon) =
\frac{g_\text{s} g_\text{v}|\varepsilon|}{4\pi\hbar^2 v^2_F}\biggl(1-J_2(\xi)\nonumber \\
&\qquad +\text{sgn}(\varepsilon) \mathcal{F}\bigl[J_3(\xi) - J_1(\xi) \bigr] \nonumber \\
&\qquad + \tfrac{1}{2}\mathcal{F}^2\bigl[J_2(\xi) -
J_4(\xi)\bigr]+{\cal O}({\cal F}^{3}) \biggr),
\end{align}
\end{subequations}
with $\xi=2x|\varepsilon|/\hbar v_{\rm F}$.
Away from the edge, $\mathcal{D}^{\text{bulk}}_{\rm A}+\mathcal{D}^{\text{bulk}}_{\rm B}\rightarrow g_\text{s} g_\text{v}|\varepsilon|/2\pi\hbar^2 v^2_F$ approaches the $\pm\varepsilon$-symmetric DOS of an infinite graphene sheet. The boundary effects break this electron-hole symmetry, as a manifestation of the chiral symmetry breaking by the reczag boundary condition.

\begin{figure}
\includegraphics[width=0.9\linewidth]{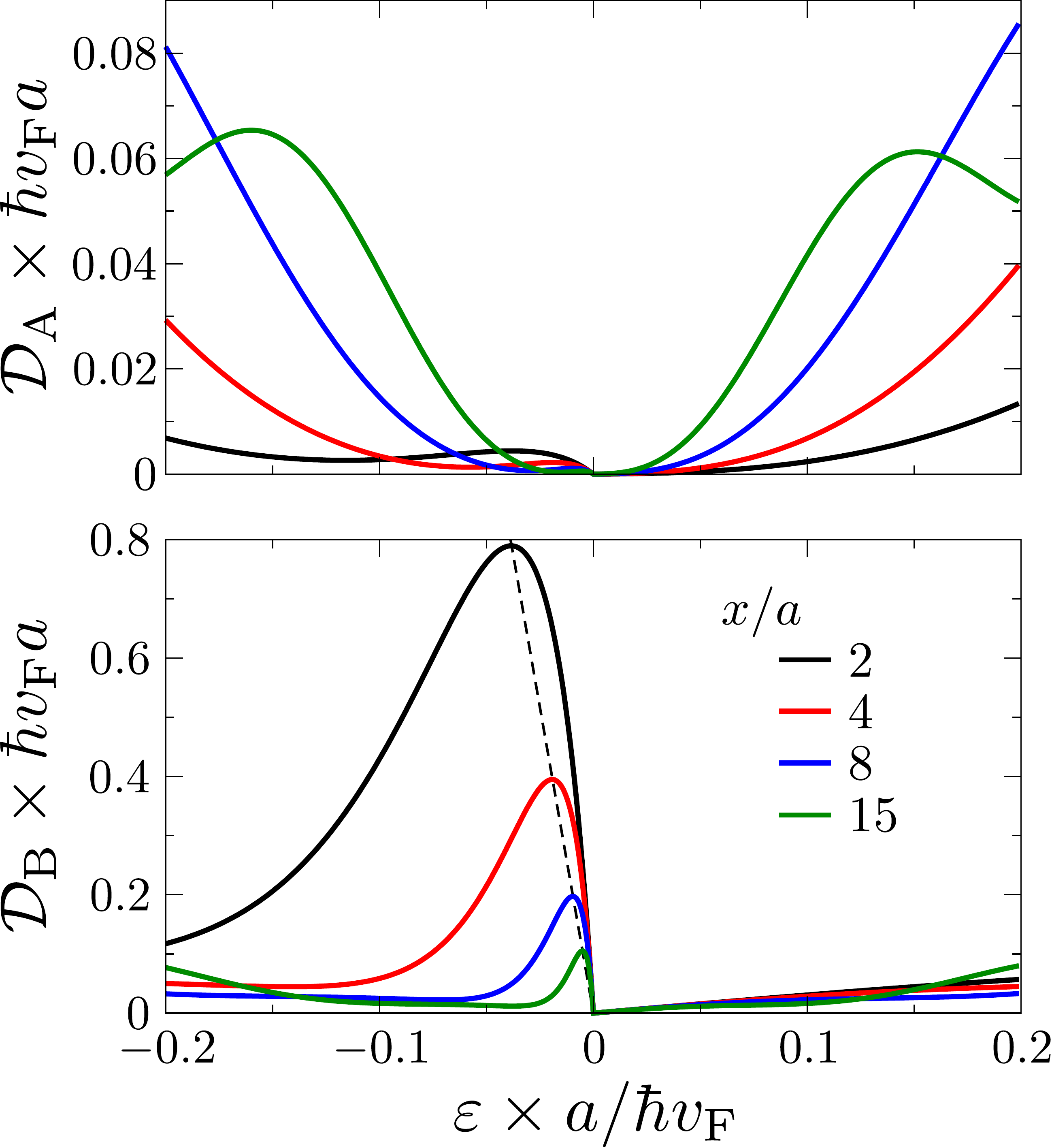}
\caption{(Color online) Local density of states as a function of energy, at various distances from the reczag edge. The contributions from the A and B sublattices are shown separately in the two panels. The peak in the density of states evolves according to Eq.\ \eqref{bc_meas} (dashed line in lower panel).}\label{edge_ldos}
\end{figure}

Fig.~\ref{edge_ldos} shows the full local DOS on each sublattice, $\mathcal{D}_{\rm X}= \mathcal{D}_{\rm X}^\text{edge}
+\mathcal{D}_{\rm X}^\text{bulk}$ with ${\rm X}\in\{{\rm A,B}\}$. 
The edge state manifests itself as a peak in the local DOS on the B sublattice. The DOS on the A sublattice is much smaller near the edge (by a factor $\mathcal{F}^2 \approx 0.006$).
The peak energy $\varepsilon_{\text{peak}}$
moves towards the Dirac point (the zero of energy) as the distance $x$ from the edge is increased,
according to
\begin{equation}\label{bc_meas}
\frac{\varepsilon_\text{peak}}{\hbar v_{F}}=\frac{\vartheta}{2x}, 
\end{equation}
for $x \gtrsim 3a/2$, $|\vartheta|\ll\pi/2$. (The Dirac approximation breaks down at smaller $x$, while for larger $\vartheta$ the edge DOS no longer dominates over the bulk DOS.)
We conclude that STM experiments have direct access to the boundary condition
angle $\vartheta$, through the dependence of
the edge state peak on the distance from the edge.

\subsection{Nanoribbon}

So far we considered a semi-infinite graphene sheet with a single B-type reczag edge. Reczag nanoribbons (width $W$) will have a B-type reczag edge on one side (at $x=0$) and an A-type reczag edge at the other side (at $y=W$). The spectrum now consists of a discrete set of transverse modes $\varepsilon_{n}(k)$, governed by the transcendental
equation\cite{Akhmerov2008a}
\begin{equation}
\begin{split}
\cos^2\vartheta \left(\cos\omega - \cos^2\Omega\right)- \sin^2\vartheta \cos\omega \cos^2\Omega\\
+\sin\Omega \left(\sin\Omega - \sin\omega \sin2 \vartheta \right) = 0.
\end{split}
\end{equation}
We defined $\omega^2=4 W^2 [(\varepsilon/\hbar v_\text{F})^2 - k^2]$ and $\cos \Omega = \hbar v_\text{F}
k/\varepsilon$.

\begin{figure}
\includegraphics[width=\linewidth]{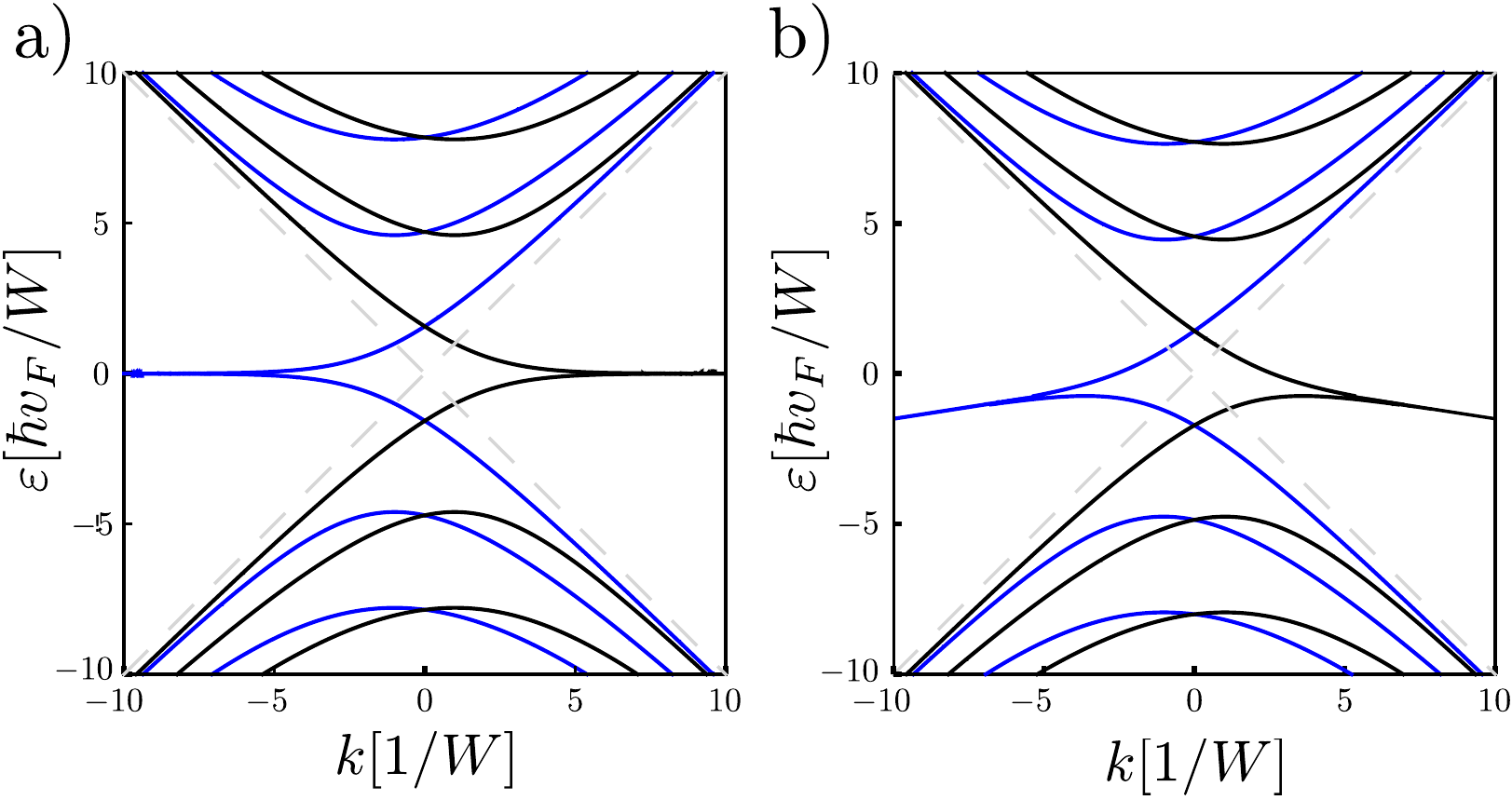}
\caption{(Color online) Comparison of the dispersion relations of modes in zigzag (a) and (b) reczag nanoribbons. The results for both valleys are superimposed, by measuring $k$ relative to the $K$-point (black curves) and $K'$-point (blue curves), respectively. The dashed grey lines
indicate the Dirac cone of graphene.}\label{nanoribbon}
\end{figure}

Fig.~\ref{nanoribbon} compares the mode dispersion of a zigzag nanoribbon\cite{Brey2006a} ($\vartheta=0$) and a reczag nanoribbon. The prominent difference is 
the dispersion of the reczag edge mode: For $kW\cos\vartheta  \gg 1$ it is given (up to exponentially
small corrections) by the results for a single edge, $\varepsilon(k)=\pm \hbar v_\text{F}\left|k\right|
\sin\vartheta $, since then the wave functions on opposite edges
decay rapidly and overlap only little. Both in the zigzag and reczag nanoribbon, the overlap of the edge states as $k \rightarrow 0$ produces a larger and larger energy splitting, until the edge states 
merge with the bulk bands. 

The bulk bands of the reczag nanoribbon have a slight offset towards towards
negative energies (barely visible in Fig.\ \ref{nanoribbon}), which breaks the electron-hole symmetry --- again as a result of the breaking of chiral symmetry by the edge reconstruction.

\section{Effect of a magnetic field}\label{sec_mag}

\subsection{Dirac solutions}

The presence of a uniform perpendicular magnetic field $B_{0}$ is accounted for by the substitution
$\bm{p} \mapsto \bm{p} + e \bm{A}$, with $-e$ the electron charge and $\bm{A} = B_{0}x\hat{y}$ the vector potential in the Landau gauge.
The valleys remain uncoupled and translational invariance along the $y$-axis is preserved.
The wave function $\Psi(x,y)=
\psi(x) e^{iky}$ in a single valley thus satisfies the Dirac equation
\begin{subequations}
\label{dirac_mag}
\begin{align}
\label{dirac_mag_1}
\psi_1 &= -i \frac{\sqrt{2}}{E}\left(\partial_X + \tfrac{1}{2}X \right)\psi_2, \\
\label{dirac_mag_2}
\psi_2 &= -i \frac{\sqrt{2}}{E}\left(\partial_X - \tfrac{1}{2}X \right)\psi_1,
\end{align}
\end{subequations}
where $E =\varepsilon l_\text{m}/\hbar v_{\rm F}$, $X = \sqrt{2}\left(x/l_\text{m} +
kl_\text{m}\right)$, and $l_\text{m}=\sqrt{\hbar/eB_{0}}$ is the magnetic length.

The coupled first-order differential equations \eqref{dirac_mag} decouple into a second-order equation,
\begin{equation}\label{eff_mag}
\partial_X^2\psi_j(x) = \left(\tfrac{1}{4}X^2 -\tfrac{1}{2}E^2 \pm \tfrac{1}{2}
\right) \psi_j(x)\,,
\end{equation}
where $j=1,2$ and the plus sign holds for $\psi_1$ while the minus sign
holds for $\psi_2$. Eq.~\eqref{eff_mag} is solved by the parabolic
cylinder function $\mathcal{U}(x,a)$, determined up to normalization by\cite{Abr72}
\begin{equation}
\partial_x^2\,\mathcal{U}=(\tfrac{1}{4} x^2+a)\mathcal{U},\;\;\lim_{x\rightarrow\infty}\mathcal{U}(a,x)=0.
\end{equation}
The solution in a magnetic field takes the form
\begin{subequations}
\label{magnetic_solution}
\begin{align}
\psi_1 &=
\frac{E}{\sqrt{2}}\left[A\, \mathcal{U}\left(\frac{1-E^2}{2},
X\right) - B\,\mathcal{U}\left(\frac{1-E^2}{2}, -X \right) \right]\,, \\
\psi_2&= i A\,\mathcal{U}\left(-\frac{1+E^2}{2}, X \right) +
iB\,\mathcal{U}\left(-\frac{1+E^2}{2}, -X\right)\,,
\end{align}
\end{subequations}
where $A$ and $B$ are constants.

\subsection{Edge states and Landau levels}

We first consider a semi-infinite graphene sheet for $x\geq 0$,
terminated by a B-type reczag edge at $x=0$. Only keeping the solutions that
decay for $x\rightarrow \infty$ in Eq.\ \eqref{magnetic_solution} and
substituting the boundary condition \eqref{bc_isotropic_B}, we obtain
an implicit equation for the energy dispersion in the two valleys,
\begin{equation}\label{mag_disp1}
\frac{E}{\sqrt{2}} = \frac{\mathcal{U}\left(-\frac{1 + E^2}{2},
\sqrt{2}kl_\text{m}
\right)}{\mathcal{U}\left(\frac{1-E^2}{2}, \sqrt{2}kl_\text{m}\right)}\times\left\{\begin{array}{ll}
-\mathcal{F}&\text{in valley}\;K,\\
1/\mathcal{F}&\text{in valley}\;K'.
\end{array}\right.
\end{equation}

\begin{figure}
\includegraphics[width=0.8\linewidth]{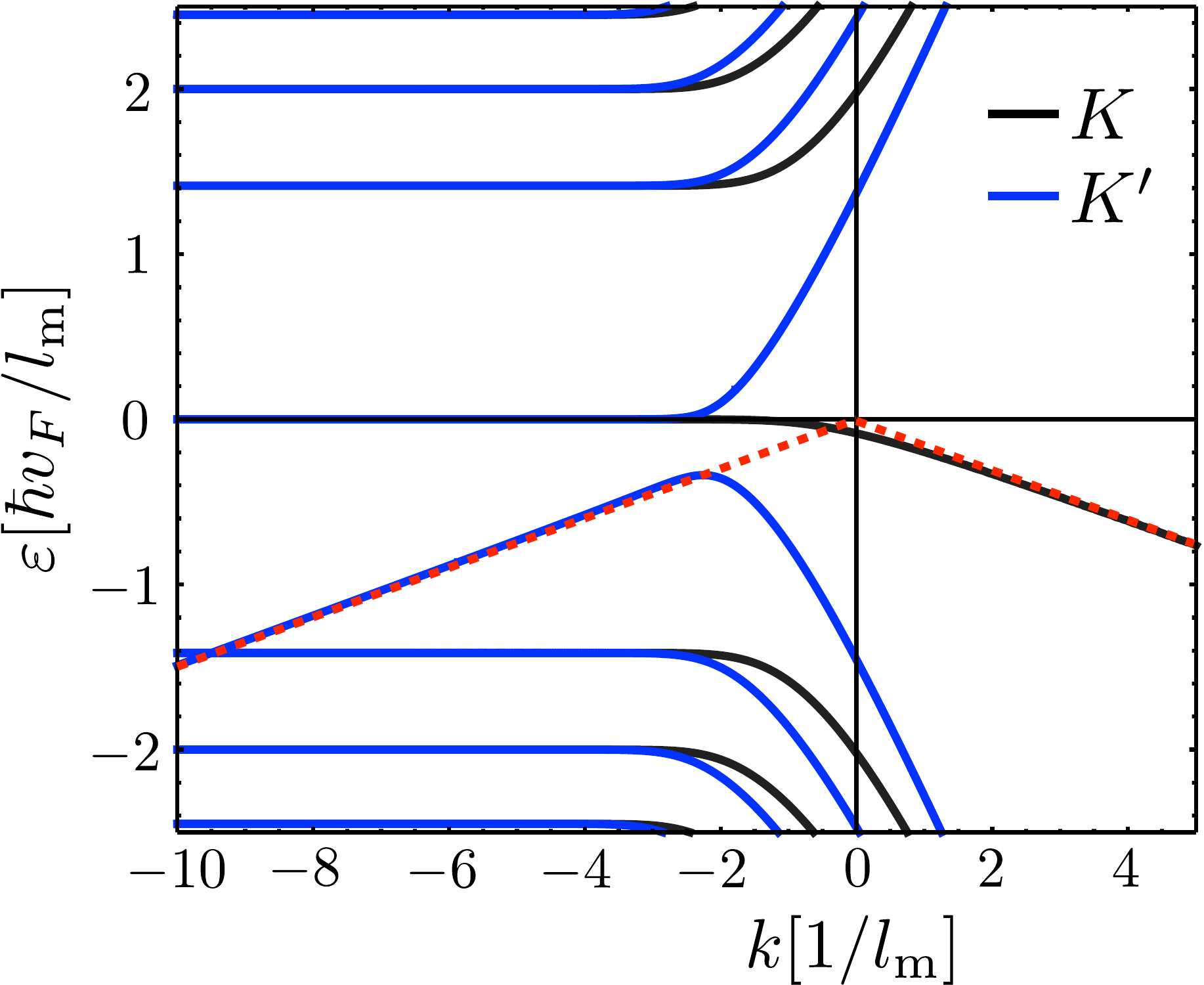}
\caption{(Color online) Energy dispersion at the B-type reczag edge in a magnetic field.
The states in valley $K$ and $K'$ are shown in black and blue, respectively
(with $k$ measured relative to the respective Dirac point). The zero-field edge
state dispersion is included as dashed, red lines.}\label{magnetic_disp}
\end{figure}

The resulting dispersion is shown in Fig.~\ref{magnetic_disp}. The main features can be
understood from two principles:
\begin{itemize}
\item
The confining potential due to
the magnetic field in Eq.~\eqref{eff_mag} has its minimum at
$-k l_\text{m}^2$. Because of this, we find bulk-like Landau level solutions
and hence flat bands for $k \ll 0$ with the bulk Landau level energy\cite{Ando2005}
$\varepsilon_n=\text{sgn}(n)(\hbar v_\text{F}/l_\text{m}) \sqrt{2 |n|}$, $n \in 
\mathbb{Z}$. For positive values of $k$, the center of the confining potential is
moved beyond the edge of the sample, resulting in dispersive quantum
Hall edge states with velocity $v_\text{F}$ (larger than the velocity $v_\text{F} \sin\vartheta$ of the zero-field reczag edge states).
\item
The magnetic field has little effect on the 
reczag edge states, if the edge state decay length is smaller than the magnetic length,
$\left| k \cos\vartheta \right|^{-1} \ll l_\text{m}$. For this reason, we observe two bands
in Fig.~\ref{magnetic_disp} that follow the reczag edge dispersion (shown as dashed lines)
for large enough momenta. 
\end{itemize}

\subsection{Triple edge mode in the lowest Landau level}

\begin{figure}
\includegraphics[width=0.9\linewidth]{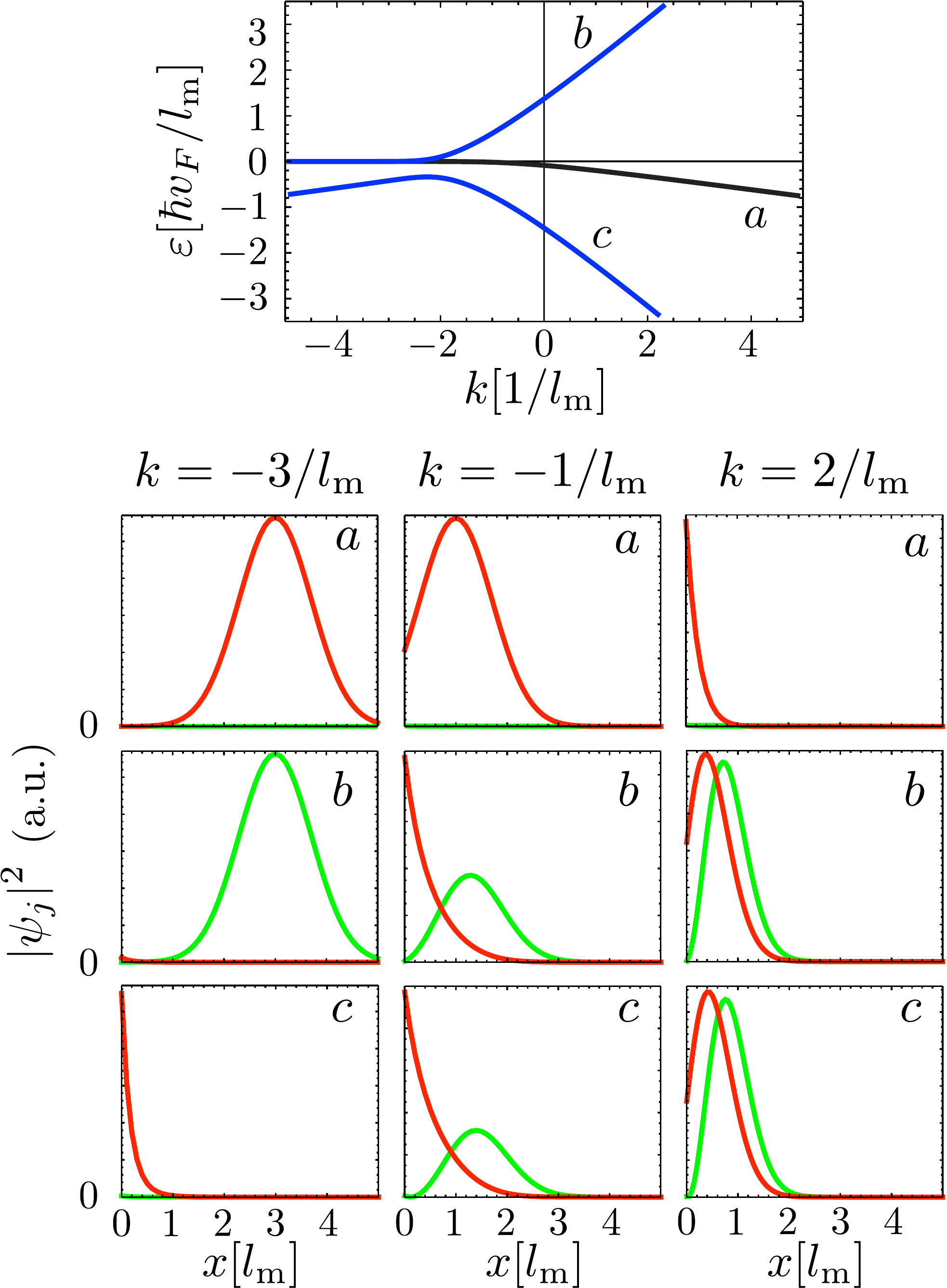}
\caption{(Color online) Nine lower panels: Probability density profiles for three different values of $k$ in the three distinct modes of the lowest Landau level, labeled $a,b,c$ in the top panel. Mode $a$ is in valley $K$, and the counter-propagating modes $b,c$ are in valley $K'$.
The colors distinguish the probability densities
on sublattice A (green) and B (red). To allow a comparison of the profiles, the vertical axis in each graph has been rescaled.}\label{magnetic_wf}
\end{figure}

The interplay of the magnetic and zero-field edge states produces three distinct edge modes in the lowest Landau level
($n=0$). These are labeled $a,b,c$ in the top panel of Fig.\ \ref{magnetic_wf}. The unidirectional edge mode $a$ in valley $K$ is accompanied by a pair of counterpropagating edge modes in valley $K'$. These three modes have a distinct wave function profile, as shown in the lower panels of Fig.\ \ref{magnetic_wf}.

For mode $a$ in the $K$-valley, the bulk Landau level solution for $k\ll 0$ is nonzero on the 
B sublattice only.\cite{Ando2005} It moves closer to the edge with increasing $k$
and eventually becomes the reczag edge state, which is mostly localized on sublattice 
B, with a small ${\cal O}(\mathcal{F}^2)$ contribution on the A sublattice.
In contrast, for modes $b,c$ in the $K'$-valley, there are two solutions for every momentum: For $k\ll 0$
we find both the bulk Landau level solution (localized on sublattice A only) and the reczag edge state
(localized mostly on sublattice B). Note that we find bands with a distinct bulk \emph{or} edge character,
in contrast to the zigzag edge where chiral symmetry forces always hybridized solutions.\cite{Brey2006}

The tripling of the edge modes in the lowest Landau level does not change the value of the Hall conductance, since the contribution from the two counterpropagating modes cancels. But the valley polarization at the edge is changed.
At a zigzag edge, the lowest Landau level edge modes are in the same valley for positive and negative energies, whereas they are in different valleys at an armchair edge.\cite{Akhmerov2007}
At the reczag edge both valleys are present for negative energy, with only a single valley for positive energy.

\subsection{Comparison with tight-binding model}

\begin{figure}
\includegraphics[width=\linewidth]{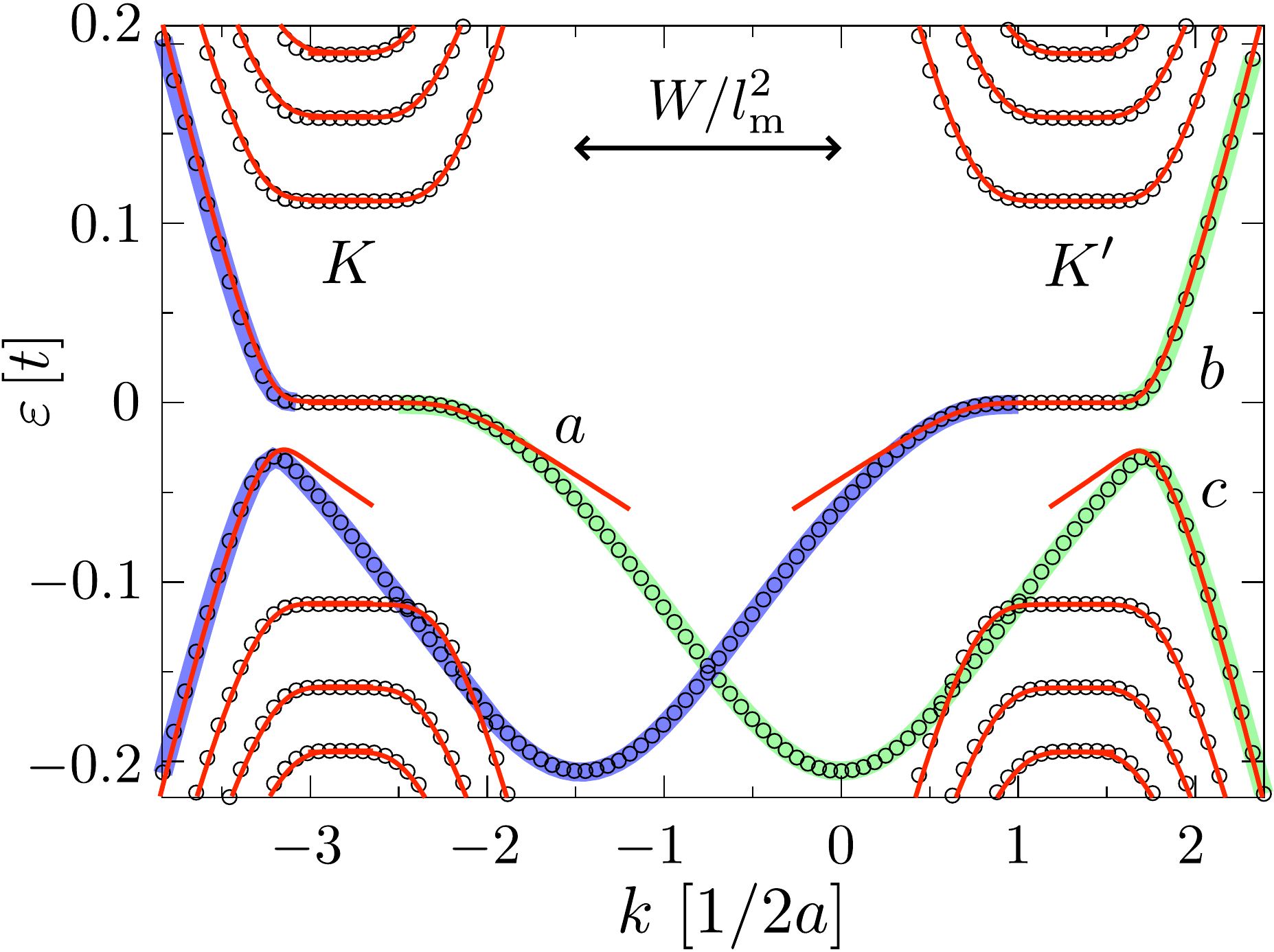}
\caption{(Color online) Comparison of the energy dispersion of a reczag nanoribbon in a magnetic field obtained
from the tight-binding model (open circles) and from  the Dirac equation (red lines).
For the lowest Landau level, the
edge states localized at the $x=0$ and $x=W$ boundary are highlighted in
green and blue, respectively. The three lowest-Landau-level modes at the $x=0$ edge are labeled $a,b,c$ . They appear displaced relative to Fig.\ \ref{magnetic_wf}, because there the momentum $k$ is measured relative to the Dirac point of valley $K,K'$.
}
\label{mag_ribbon}
\end{figure}

Fig.~\ref{mag_ribbon} shows a comparison between the band structure obtained from the Dirac
equation and from the tight-binding model. (Similar tight-binding calculations are in Refs.~\onlinecite{Rakyta2010,Rodrigues2011}.)
To be able to identify the contributions from the two edges we took a wide nanoribbon, $W=8\,l_\text{m} = 101 \sqrt{3}/2 a$, in which opposite edges are approximately decoupled.
In this case the Dirac equation results for the A-type reczag edge at $x=W$ can be directly
obtained from the results for a B-type reczag edge at $x=0$ by interchanging the valleys and replacing 
$k \rightarrow -k-W/l_\text{m}^2$.

The two calculations agree very well near the Dirac points. As in the zero-field case (Fig.\ \ref{comp_tb_dirac}b)
the tight-binding model connects the edge states from the two valleys $K,K'$, which are disconnected in the Dirac equation.

\section{Conclusion}\label{concl}

In conclusion, we have derived the boundary condition for the Dirac equation at reconstructed zigzag edges in graphene.
The $\vartheta$-class of boundary conditions \eqref{special_bc} applies to reconstructions  with a unit cell that is not a multiple of three times the zigzag unit cell.
We have calculated the angular parameter $\vartheta$ for the zz(57) (reczag) reconstruction,
which has been
identified as the most stable reconstruction. Most of our results are given for general $|\vartheta|<\pi/2$, so they apply to other reconstructions in the $\vartheta$-class as well.

The $\vartheta$-class reconstructions share two key properties: they do not cause intervalley scattering and they support edge states.
Dispersive edge states were previously found for the 
reczag edge,\cite{Rodrigues2011} the zigzag edge with next-nearest neighbor
hopping,\cite{Sasaki2006} and the zigzag edge with a boundary
potential.\cite{Bhowmick2010} Our analysis identifies an entire class of reconstructions with edge states, and gives analytic expressions for the edge state dispersion in terms of a single parameter $\vartheta$.

The edge mode appears in the local density of states as a peak at energy $\varepsilon_{\rm peak}$. The dependence of $\varepsilon_{\rm peak}$ on the separation $x$ from the edge, given by Eq.\ \eqref{bc_meas}, allows a direct measurement of $\vartheta$ by scanning tunneling microscopy. 

In a magnetic field there appears a tripling of the edge modes in the lowest Landau level. This could be observed in transport experiments, since two of three edge modes are counterpropagating and therefore susceptible to localization by disorder. With increasing disorder, the two-terminal conductance would then be reduced by a factor $1/3$.

\begin{acknowledgments}
We thank A. Fasolino for drawing our attention to this problem.
Our research was supported by the Dutch Science
Foundation NWO/FOM, by the Eurocores program EuroGraphene,
and by an ERC Advanced Investigator grant.
\end{acknowledgments}

\begin{appendix}

\section{Condition for absence of valley mixing by edge reconstruction}\label{app_zonefolding}

We explain the zone-folding argument used in Sec.\ \ref{tbmodel_sec} to identify
which periodicity of the edge reconstruction
leaves the valleys uncoupled. It is similar to the 
zone-folding argument that distinguishes metallic and semiconducting 
carbon nanotubes.\cite{Saito2003}

The projection of the $K$-point along the direction of the edge is given
by
\begin{equation}
\bm{K}\cdot\frac{\bm{T}}{\left|\bm{T}\right|}= \frac{1}{3} (n-m) \frac{2\pi}{\left|\bm{T}\right|}\,,
\end{equation}
and the projection of the $K'$-point by
\begin{equation}
\bm{K'}\cdot\frac{\bm{T}}{\left|\bm{T}\right|}= \frac{1}{3} (m-n) \frac{2\pi}{\left|\bm{T}\right|}\,.
\end{equation}
The projected $K$ and $K'$-points correspond to the same momentum in the 
one-dimensional first Brillouin zone of the edge, if they differ by a multiple of a reciprocal
lattice vector. This condition $(\bm{K}-\bm{K'})\cdot \bm{T}/\left|\bm{T}\right| = 
l\, 2\pi/\left|\bm{T}\right|$, $l \in \mathbb{Z}$, is equivalent to
the condition that $n-m$ is divisible by 3. Otherwise, if $n\neq m$ mod $3$, the $K$-points project to different momenta in the 
first Brillouin zone of the edge, and since these momenta are conserved due to
translational symmetry, the valleys remain uncoupled.

\section{Boundary condition for modified zigzag edge}\label{app_modified_zigzag}

Edge reconstruction is one modification of the zigzag edge that leads to a boundary condition of the single-parameter form \eqref{special_bc}.
In this Appendix we calculate the value of the parameter $\vartheta$ for two alternative modifications of the zigzag edge that break chiral symmetry: On-site potentials and next-nearest-neighbor hopping. Since most of our results for the reczag edge are given for arbitrary $\vartheta$, they can be applied to these edges as well --- even though these modifications leave the lattice structure unaffected.

\begin{figure}
\includegraphics[width=\linewidth]{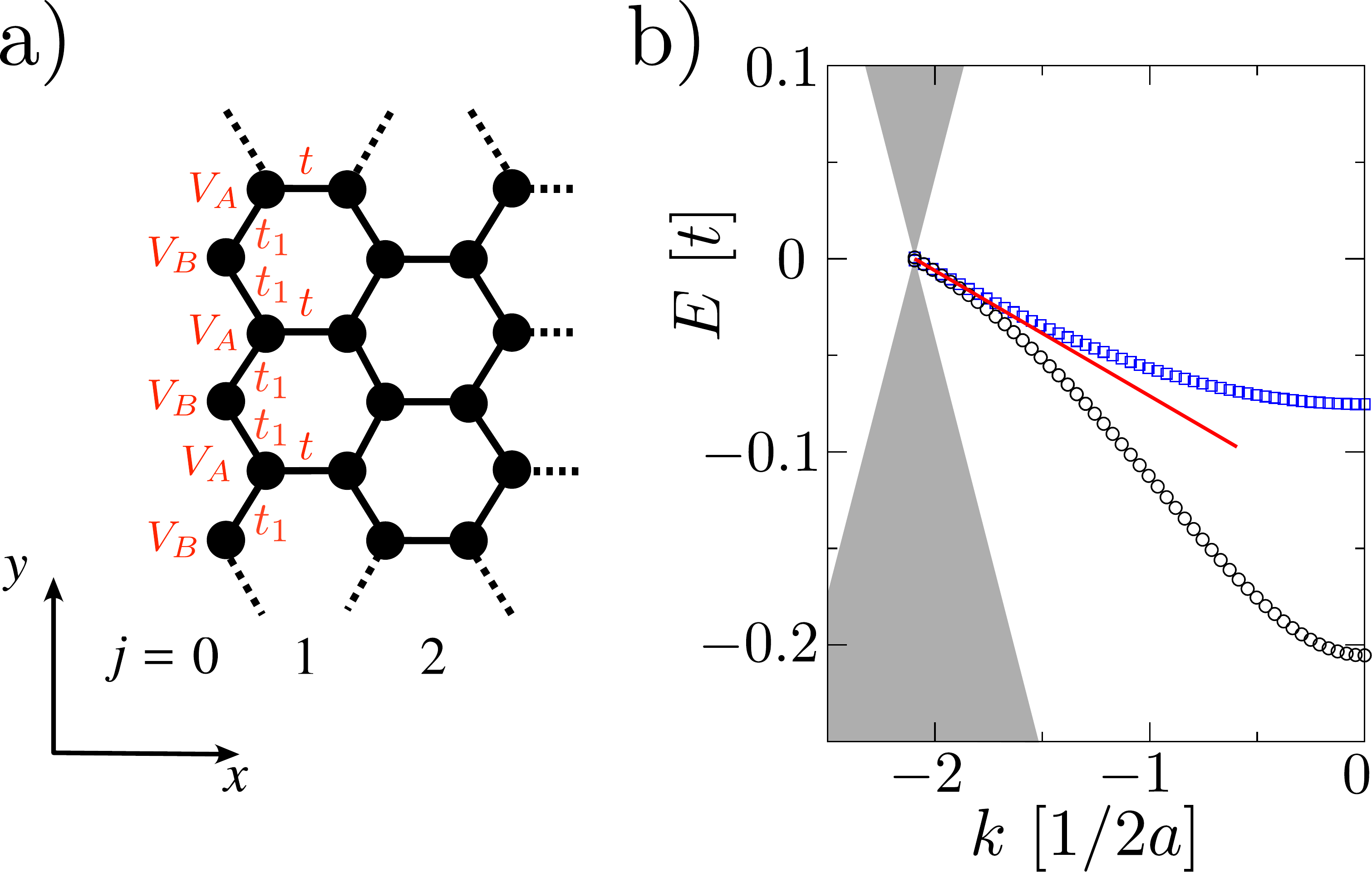}
\caption{(Color online) (a) Schematic of the modified zigzag edge, with 
on-site potentials and hoppings 
labeled in red. (b) Comparison between the tight-binding edge state
dispersion for the reczag edge (black circles), and the 
modified zigzag edge with $V_{\rm A}=0$, $t_1=t$, $V_\text{B}=-\mathcal{F} t$
(blue squares). The Dirac equation has the same boundary condition at these two edges, leading to the same 
edge state dispersion near the Dirac point (red solid
line).
}\label{mod_zigzag}
\end{figure}

Consider a B-type zigzag edge with 
a nonzero potential $V_\text{A}$, $V_\text{B}$ on the outermost 
A and B atoms. (See Fig.~\ref{mod_zigzag}a.) Such on-site potentials could appear because the edge atoms see a different chemical environment
than the bulk atoms. We also include a possible
modification $t_{1}$ of the hopping amplitude at the edge.
The same model with $V_\text{B}=-t'$ describes to leading order the effect of
a next-nearest-neighbor hopping $t'$.\cite{Sasaki2009}

Since the unit cell is not changed by these modifications, the boundary modes
that appeared for the reczag edge are absent. 
Following the approach of Sec.~\ref{sec_reczag_bc} we find
\begin{equation}\label{mod_zigzag_F}
\mathcal{F}=\tan(\vartheta/2) = \frac{tV_\text{B}}{V_\text{A}V_\text{B}-t_1^2}\,.
\end{equation}
This agrees with Refs.~\onlinecite{Wurm2011,Bhowmick2010} for the special case $V_\text{A}=0$, $t_1=t$.
If next-nearest-neighbor hopping is the only modification, we set 
$V_\text{B}=-t'$, $V_\text{A}=0$, $t_1=t$ and arrive at
\begin{equation}
\mathcal{F}=\tan(\vartheta/2) = t'/t\,.
\end{equation}

Fig.~\ref{mod_zigzag}b
shows a comparison of the edge state dispersion for the
reczag edge from Sec.~\ref{sec_reczag_bc} and a
zigzag edge with an edge potential such that the value of ${\cal F}$ is the same.
Both have the same boundary condition for the Dirac equation, and
indeed we observe the same linearly dispersing edge state close to the 
Dirac point.

\section{Extended model for the reczag edge}\label{app_extended_reczag}

The tight-binding model for the reczag edge used in the main text 
is based on Ref.~\onlinecite{Rakyta2010}. An extended
model was studied in Ref.~\onlinecite{Rodrigues2011}, including
also modifications of the hopping amplitudes in the first
row of hexagons near the edge. From the general
arguments of Sec.~\ref{sec_general_bc} we know that the form of the boundary
condition remains the same, with a different numerical value for the parameter
$\vartheta$. In this Appendix we calculate that value.

\begin{figure}[tb]
\includegraphics[width=0.6\linewidth]{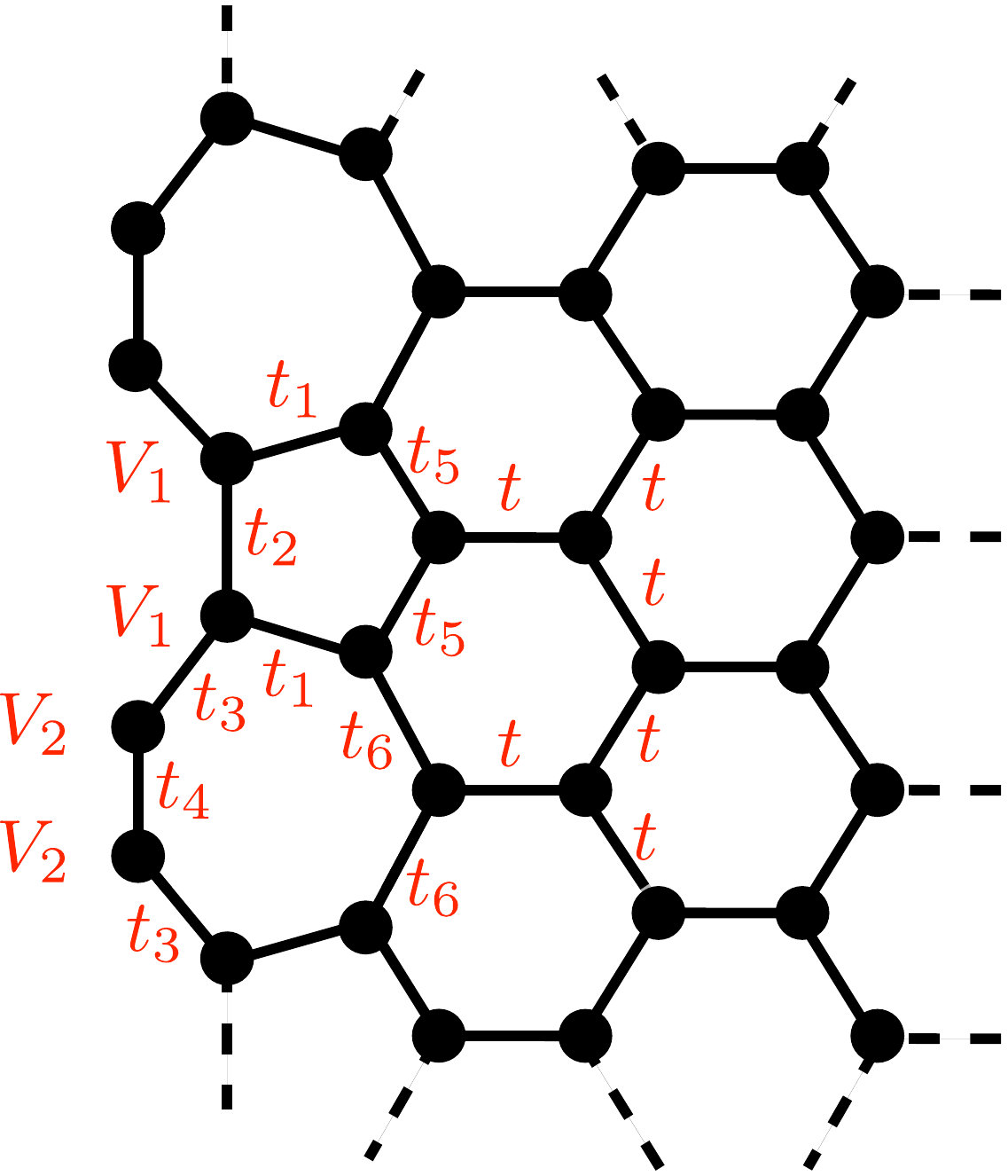}
\caption{(Color online) Schematic of the extended model for the
reczag edge, with on-site energies and
hoppings labeled in red.}\label{extended_reczag}
\end{figure}

The extended model of the reczag edge is shown in Fig.~\ref{extended_reczag}.
In addition to the modified hopping amplitudes of 
Ref.~\onlinecite{Rodrigues2011}, we also include (for additional generality) an on-site
potential at the outermost edge atoms. 
Following the same procedure as in Sec.~\ref{sec_reczag_bc}, we obtain
\begin{equation}
 \mathcal{F} =\tan(\vartheta/2) = \mathcal{T}/\mathcal{N},\label{FTN}
\end{equation}
as the ratio of the coefficients 
\begin{widetext}
\begin{align}
\mathcal{T} ={}& t t_1^2\left[\left\{t_2 (2 t_5^2 + 2 t_5t_6 - t_6^2) - 
2 (t_5^2 + t_5t_6 + t_6^2)V_1\right\} (t_4^2 - V_2^2)\right.\nonumber\\
&\left.+t_3^2 \left\{t_4
(t_5^2 - 2 t_5 t_6 - 2 t_6^2) - 2 (t_5^2 + t_5 t_6 + t_6^2) V_2\right\}\right],\\
\mathcal{N} ={}& 6t_5^2t_6^2\left[t_3^4 + (t_2^2 - V_1^2)(t_4^2 - V_2^2) + t_3^2(t_2t_4 - 2 V_1V_2)\right]. 
\end{align}
\end{widetext}

\begin{table}[!tb]
\begin{tabular}{|c|c|c|c|c|c|c|c|}
\hline
$t_1/t$&$t_2/t$&$t_3/t$&$t_4/t$&$t_5/t$&$t_6/t$&${\cal F}$&$\vartheta$\\
\hline\hline
$0.94$&$0.94$&$1.06 $&
$1.42$&$1.04$&$0.98$&0.0485&0.0968\\ \hline
\end{tabular}
\caption{Values of the hopping amplitudes in the extended tight-binding
model, obtained from DFT.\cite{Rodrigues2011} In this model, $V_1=V_2=0$.
The boundary condition parameter is calculated from Eq.\ \eqref{FTN}.
}
\label{tb_params2}
\end{table}

Using the numerical values from Ref.~\onlinecite{Rodrigues2011}, see Table \ref{tb_params2}, we find
$\vartheta \approx  0.0968$ --- within a factor of two from the value $\vartheta\approx 0.150$ following from the simpler model of Table \ref{tb_params}.

\end{appendix}

\end{document}